\documentclass[a4paper,UKenglish,cleveref, autoref, thm-restate, 
]{lipics-v2021}

\usepackage{hyperref}
\usepackage{amsfonts,amsmath,amsthm,amssymb}
\usepackage{mathrsfs}  
\usepackage{cancel}
\usepackage{mathtools}
\usepackage{thm-restate} 
\usepackage{verbatim} 
\usepackage{tikz}
\usepackage{quantikz}
\usepackage{url}
\usepackage[T1]{fontenc}
\usepackage{graphics}

\def\pp{\mathinner{\ldotp\ldotp}}	

\author{Simone {Faro}}{Universit\`a di Catania, Dipartimento di Matematica e Informatica, Catania, Italia \and \url{http://www.dmi.unict.it/faro}}{simone.faro@unict.it}{https://orcid.org/0000-0001-5937-5796}{Supported by ICSC – Centro Nazionale di Ricerca in High-Performance Computing, Big Data and Quantum Computing; and by University of Catania, progetto PIACERI 2024-2026 - Linea di Intervento 1.}

\author{Arianna {Pavone}}{Department of Mathematics and Computer Science, University of Palermo, via Archirafi n.34, 90123, Palermo, Italy}{ariannamaria.pavone@unipa.it}{0000-0002-8840-6157}{Supported by PNRR project ITSERR - Italian Strengthening of the ESFRI RI RESILIENCE}

\author{Caterina {Viola}}{Universit\`a di Catania, Dipartimento di Matematica e Informatica, Catania, Italia \and \url{http://www.dmi.unict.it/viola}}{caterina.viola@unict.it}{https://orcid.org/0000-0002-7312-5002}{Supported by ICSC – Centro Nazionale di Ricerca in High-Performance Computing, Big Data and Quantum Computing.}

\authorrunning{S.Faro, A.Pavone and C.Viola}

\Copyright{Simone Faro, Arianna Pavone and Caterina Viola}


\EventEditors{John Q. Open and Joan R. Access}
\EventNoEds{2}
\EventLongTitle{42nd Conference on Very Important Topics (CVIT 2016)}
\EventShortTitle{CVIT 2016}
\EventAcronym{CVIT}
\EventYear{2016}
\EventDate{December 24--27, 2016}
\EventLocation{Little Whinging, United Kingdom}
\EventLogo{}
\SeriesVolume{42}
\ArticleNo{23}

\newcommand{\ECLOSE}{\mathrm{ECLOSE}\xspace}

\renewcommand{\t}{T\xspace}
\newcommand{\p}{P\xspace}
\def\pp{\mathinner{\ldotp\ldotp}}

\newcommand{\hor}{Horspool\xspace}
\newcommand{\quick}{Quick-Search\xspace}

\newcommand{\OR}{\ | \ }

\newcommand{\cross}{\textsc{Cross-Sampling}\xspace}

\newcommand{\bpcross}{\textsc{BP-Cross-Sampling}\xspace}

\DeclareMathOperator{\bigO}{\mathcal{O}}

\nolinenumbers

\title{Bridging Classical and Quantum String Matching: A Computational Reformulation of Bit-Parallelism}
\titlerunning{A Computational Reformulation of Bit-Parallelism}
\begin{document}

\maketitle
\begin{abstract}
\emph{String matching} is a fundamental problem in computer science, with critical applications in text retrieval, bioinformatics, and data analysis. Among the numerous solutions that have emerged for this problem in recent decades, \emph{bit-parallelism} has significantly enhanced their practical efficiency, leading to the development of several optimized approaches for both exact and approximate string matching.  
However, their potential in \emph{quantum computing} remains largely unexplored. This paper presents a novel pathway that not only translates bit-parallel string matching algorithms into the quantum framework but also enhances their performance to achieve a quadratic speedup through Grover’s search.  
By embedding quantum search within a bit-parallel model, we reduce the time complexity of string matching to \( \tilde{\mathcal{O}}(\sqrt{n}) \), establishing a structured pathway for transforming classical algorithms into quantum solutions with provable computational advantages. Beyond exact matching, this technique offers a foundation for tackling a wide range of non-standard string matching problems, opening new avenues for efficient text searching in the quantum era.  
To demonstrate the simplicity and adaptability of the technique presented in this paper, we apply this translation and adaptation process to two landmark bit-parallel algorithms: Shift-And for exact pattern matching and Shift-Add for approximate string matching with up to \( k \) errors.


\end{abstract}

\newpage

\section{Introduction}
String matching is a fundamental problem in computer science, playing a key role in text processing, information retrieval, and computational biology. It involves identifying exact or approximate occurrences of a string within a text and serves as a core component in various software and operating systems.  
The importance of efficient string matching algorithms arises from the widespread use of text as a primary medium for data exchange. In large-scale text analysis, these algorithms are essential for processing vast corpora. In computer science, they facilitate data retrieval in linear file structures, while in molecular biology, they enable sequence analysis by identifying patterns in nucleotide and amino acid sequences.  


Formally, given a pattern \( x \) of length \( m \) and a text \( y \) of length \( n \), the string matching problem consists of the following tasks:  
(1) determining whether \( x \) appears in \( y \);  
(2) counting the total number of occurrences;  
(3) identifying the positions of all occurrences.  
In the case of approximate matching, an additional objective is:  
(4) identifying the pattern configurations for each occurrence in the text.  
%

In classical computation, the exact matching problem has a time complexity of \( \Omega(n) \) for problems (1), (2) and (3), while the aprroximate case can have different solutions depending on the allowed approximations. 
Among the numerous solutions available in the literature~\cite{FL13}, one of the most significant advancements in practical text searching was the introduction, in the early 1990s, of bit-parallelism for simulating non-deterministic finite-state automata~\cite{BYG92}. By making use of bitwise operations within a word RAM model, this approach reduces the time complexity of the search phase by a factor proportional to the word size.  
This technique has inspired extensive research in both standard and non-standard string matching~\cite{FL13}.

Three decades later, quantum computing is once again having a significant impact on text processing \cite{GallS23,LCS24} and text searching \cite{RAMESH2003103,Montanaro14,Niroula21,QPP24}. The first quantum solution to exact string matching, developed by Ramesh and Vinay~\cite{RAMESH2003103}, combines Grover's search~\cite{Grover96} with a parallel string matching technique. Their algorithm operates in the quantum query model~\cite{Cleve_2001} and addresses problem (1) with \(\tilde{\mathcal{O}}(\sqrt{n})\) queries. Thus, problems (2) and (3) can be solved in \(\tilde{\mathcal{O}}(\sqrt{\rho n})\) time,\footnote{Whenever the number of solutions is \(\rho>1\), and we would like to find all of them, \(\Theta(\sqrt{\rho n})\) iterations of Grover's search are sufficient and necessary.} where \(\rho\) is the number of occurrences of \( x \) in \( y \), providing a real quantum advantage when \(\rho=o(n)\). For the average case, Montanaro~\cite{Montanaro14} demonstrated a super-polynomial separation between quantum and classical complexity.
The first solution in the quantum circuit model~\cite{Yao93} with a \(\mathcal{\tilde{\mathcal{O}}}(\sqrt{n})\)-time complexity for (1) was presented by Niroula and Nam in 2021~\cite{Niroula21}, then extended~\cite{CFP23} to swap matching with a \(\mathcal{\tilde{\mathcal{O}}}(\sqrt{\rho n2^m})\) complexity for (4).


In this paper, we demonstrate how a classical bit-parallelism-based algorithm with linear complexity can be transformed into a quantum algorithm that solves the same problem in \( \tilde{\mathcal{O}}(\sqrt{n}) \) time. While this result does not improve the complexity of the standard exact string matching problem—matching the result of Ramesh and Vinay~\cite{RAMESH2003103}—it introduces a more general framework that extends to any bit-parallel text searching algorithm, including those designed for non-standard matching problems.  

To showcase the effectiveness of our approach, we apply it to two of the most well-known bit-parallel algorithms: Shift-And for exact matching and Shift-Add for approximate matching with up to \( k \) errors. This technique paves the way for translating a wide range of bit-parallelism-based algorithms into the quantum domain, enabling quantum speedups in specific scenarios.

\section{String Matching and Bit-Parallelism}\label{sect:single}

Finite-state automata play a central role in both standard and non-standard string matching, enabling efficient text scanning strategies. The Knuth-Morris-Pratt algorithm, for example, recognizes substrings of the pattern using a deterministic automaton, achieving worst-case linear time complexity. Conversely, the Backward-DAWG-Matching algorithm processes text substrings in reverse, attaining optimal average-case complexity.  

While deterministic automata provide strong theoretical guarantees, non-deterministic automata offer a more compact and efficient representation for string matching, though their simulation in the word-RAM model is generally inefficient.  
To address this, \emph{bit-parallelism} was introduced~\cite{Dom68} and later refined~\cite{BYG92,WM92} to efficiently simulate non-deterministic finite-state automata. Using the parallelism of bitwise operations, this technique reduces the number of operations required by a factor of up to \( \omega \), where \( \omega \) is the word size of the computer.

Algorithms designed in the bit-parallelism framework represent each state of the automaton using a single bit within a memory register so that the entire automaton can be represented by the bit vector stored in a computer word, provided that the number of states does not exceed its size.
Then, they use bitwise operations, i.e., operations that manipulate one or more bit vectors at the level of their individual bits, to simulate the automaton's transitions in parallel across all states. This allows these algorithms to reduce their execution time by a factor of $w$, where $w$ is the size of a computer word. 

On modern architectures bitwise operations have the same speed as addition but are significantly faster than multiplication and division.
In this paper we use the C-like notations in order to represent bitwise operations.
In particular, for the purpose of our discussion, we mention the following bitwise operations:
``$|$'' represents the bitwise operation \textsc{Or}
 ($\texttt{01101101}\ |\ \texttt{10101100} = \texttt{11101101}$); 
``$\&$'' represents the bitwise operation \textsc{And}
 ($\texttt{01101101}\ \&\ \texttt{10101100} = \texttt{00101100}$); and 
``$\ll$''  represents the bitwise left shift
 ($\texttt{01101101} \ll 2 = \texttt{10110100}$).
%
These operations are executed within a single CPU cycle.

\smallskip

The Shift-And algorithm \cite{BYG92} solves the exact string matching problem using bit-parallelism to simulate the behavior of a non-deterministic string matching automaton (NFA) that recognizes the language \(\Sigma^{*}x\) for a the input pattern \(x\) of length \(m\).  


Formally, this NFA is defined as the 5-tuple \(\text{NSMA}(x) = (Q, \Sigma, \delta, q_0, F)\). The set of states of the automaton is given by \( Q = \{q_0, q_1, \dots, q_m\} \), where \( q_0 \) represents the initial state and \( q_m \) is the unique final state, i.e., \( F = \{q_m\} \). The automaton processes symbols from the input alphabet \( \Sigma \), and its behavior is governed by the transition function \( \delta: Q \times \Sigma \to Q \). 

The structure of the automaton consists of a linear sequence of transitions that advance through its states. Specifically, for each position \( i \) in the pattern, the transition function is defined as \( \delta(q_i, x[i]) = q_{i+1} \), ensuring that the automaton progresses deterministically when reading a matching character. In addition to these transitions, a self-loop is defined at the initial state, such that \( \delta(q_0, c) = q_0 \) for all \( c \in \Sigma \), introducing non-determinism and allowing the automaton to start matching the pattern at any position in the input text.

The bit-parallel representation of the automaton is implemented using an array \( b \) of \(|\Sigma|\) bit-vectors, each of length \( m \). The \( i \)-th bit of \( b[c] \) is set if and only if the transition function satisfies \(\delta(q_i, c) = q_{i+1}\), or equivalently, if \( x[i] = c \), for \( c \in \Sigma \) and \( 0 \leq i < m \). 

The configuration of the automaton at any given step is stored in a bit-vector \( d \) of \( m \) bits. Since the initial state \( q_0 \) is always active, it does not need to be explicitly represented. The \( i \)-th bit of \( d \) is set if and only if the corresponding state \( q_{i+1} \) is active, for \( i = 0, \dots, m-1 \).  
At the beginning of the search, the bit-vector \( d \) is initialized to \( 0^m \), indicating that no states beyond \( q_0 \) are initially active.
As the text is processed sequentially from left to right, state transitions in the automaton are efficiently computed using bitwise operations. Specifically, given a current configuration \( d \), a transition on an input character \( c \) is performed using the bitwise update   
$d \leftarrow ((d \ll 1) \OR 1) \ \& \ b[c]$, 
where the bitwise OR with \( 1 \) (represented as \( 0^{m-1}1 \)) ensures that the self-loop at the initial state \( q_0 \) is correctly handled.  
If, after processing the character at position \( j \), the final state \( q_m \) is active, a match is reported at position \( j - m + 1 \).  

If the pattern length does not exceed the word size, the automaton fits within a single register, allowing the algorithm to run in linear time. Otherwise, each transition requires \( \lceil m / w \rceil \) operations, leading to a time complexity of \( \Theta(n \lceil m / w \rceil) \).

\smallskip

The Shift-Add algorithm \cite{BYG92} is a natural generalization of the Shift-And algorithm to the problem of approximate string matching with at most \( k \) errors. More formally, a pattern \( x \) occurs at position \( j \) in a text \( y \) if the substrings \( x \) and \( y[j \dots j+m-1] \) differ in at most \( k \) positions. 
%
As a consequence, in contrast to Shift-And, the Shift-Add algorithm requires \(\log m\) bits per position to track the number of mismatches encountered so far. This additional information enables the algorithm to determine whether the number of errors remains within the allowed threshold \( k \).


Given a configuration \( d \) that represents the current state of the automaton, a transition on an input character \( c \) is computed by adding the corresponding bit-vector \( b[c] \) to the shifted state vector. Fomally we perform the bitwise operation   
$d \leftarrow (d \ll 1) + b[c]$. 
This addition allows the algorithm to track mismatches, as each entry in \( d \) accumulates the number of errors encountered up to that position.  
At the beginning of the search, \( d \) is initialized to zero. As the text is processed from left to right, \( d \) is updated at each step. A match is detected at position \( j - m + 1 \) if, after \( m \) comparisons, the final entry of \( d \) contains a value less than or equal to \( k \). That is, a match occurs if $d[m-1] \leq k$.

Similar to the Shift-And algorithm, the Shift-Add algorithm runs in \( O(n \lceil m / \omega \rceil) \) worst-case time and requires \( O(\sigma \lceil m \log m / \omega \rceil) \) extra space, where \( \omega \) is the word size of the machine.


\section{Bitwise Operations: A Bridge from Classical to Quantum Computing}

Quantum computing makes use of quantum mechanics to build powerful computing systems. Unlike classical computers using binary bits (0 or 1), quantum computers use qubits, which can exist in multiple states simultaneously. This superposition, along with entanglement—where qubits perform coordinated operations—sets quantum computing apart.

Before delving into the details of our implementation, we need to briefly introduce the fundamental principles of quantum computing and the quantum model adopted in this work.




\smallskip

The fundamental unit in quantum computation is the \emph{qubit}. A qubit is a coherent superposition of the two orthonormal computational basis states, denoted by $\ket{0}$ and $\ket{1}$, using the conventional \emph{bra–ket} notation. Formally, a single qubit is an element from the \emph{state space} $\mathcal{H}$, which is the two-dimensional Hilbert space over the complex numbers equipped with an inner product. Therefore, the mathematical expression of a qubit $\ket{\psi}$ is a linear combination of the two basis states, i.e., $\ket{\psi} = \alpha \ket{0} + \beta \ket{1}$. The values $\alpha$ and $\beta$, called \emph{amplitudes}, are complex numbers such that $|\alpha|^2 + |\beta|^2 = 1$. These values represent the probabilities of measuring the qubit in the state $\ket{0}$ or $\ket{1}$, respectively. A \emph{quantum measurement} is the only operation that reveals information about the state of a qubit; however, this operation causes the qubit to collapse to one of the two basis states.
When multiple qubits are considered together, they form a \emph{quantum register}. A quantum register $\ket{\psi} = \ket{q_0, q_1, \ldots, q_{n-1}}$ of size $n$ is an element from the tensor product of $n$ state spaces, $\mathcal{H}^{\otimes n}$. It is expressed as a linear combination of the $2^n$ states in $\{0,1\}^n$, i.e., $\ket{\psi} = \sum_{k=0}^{2^n-1} \alpha_k \ket{k}$, where the values $\alpha_k$ represent the probabilities of measuring the register in the state $\ket{k}$, and $\sum_{k=0}^{2^n-1} |\alpha_k|^2 = 1$.
Let $k$ be an integer that can be represented by a binary string of length $n$. The symbol $\ket{k}$ denotes the register of size $n$ such that $\ket{k} = \bigotimes_{i=0}^{n-1} \ket{k_i}$, where $\ket{k_i}$ is the $i$-th least significant binary digit of $k$. For example, the quantum register $\ket{9}$ with 5 qubits is given by $\ket{9} = \ket{0} \otimes \ket{1} \otimes \ket{0} \otimes \ket{0} \otimes \ket{1}$.
We use $\ket{q}^{\oplus n}$ to denote a quantum register of size $n$ where each qubit is in the state $\ket{q}$.

\smallskip

In this paper, we adopt the \emph{circuit model of computation}~\cite{Cleve_2001}, introduced by Deutsch~\cite{Deutsch1989} as the \emph{quantum network} model. A quantum circuit consists of qubits moving linearly through a sequence of quantum gates, with inputs on the left and outputs on the right. At each time step, a qubit interacts with at most one gate.
The quantum circuit model extends its classical counterpart by leveraging superposition, entanglement, and parallelism~\cite{MooreN01,GreenHMP02}. While classical computation applies one gate at a time, quantum gates can operate simultaneously on different qubits. Given \( n \) qubits, up to \( n \) operations can occur per time step, enabling potential speedups analogous to a classical system with \( n \) processors.
The \emph{size} of a quantum circuit, defined by the number of gates, measures computational complexity. However, circuit \emph{depth}—the number of sequential gate layers—better characterizes execution time~\cite{BroadbentK09}. Depth is crucial for practical implementations, as quantum gates have finite operation times, and decoherence limits qubit longevity. Shallow circuits maximize qubit coherence and improve feasibility. Thus, in this work we adopt circuit depth as a measure of complexity.


\smallskip

\subsection{Basic Quantum Operators}

\emph{Operators} in quantum computing are mathematical entities (also referred as \emph{Gates}) that change the state of a quantum register. 
While any quantum operator can be realized for a fixed-size register, variable-size operators require the composition of elementary gates.
Quantum operators must be \emph{reversible}, meaning that their transformation can be uniquely inverted. This reversibility is ensured by unitary evolution, and when ancillary qubits are used, they must be properly managed to prevent loss of information, typically by \emph{uncomputing} intermediate states.  
In this section, we briefly list some basic components for building a quantum circuit, focusing on those relevant to this paper.

The \emph{Pauli}-X (or $X$ or NOT) gate is the quantum equivalent of the classical NOT gate. It operates on a single qubit, mapping $\ket{0}$ to $\ket{1}$ and $\ket{1}$ to $\ket{0}$.
The \emph{Hadamard} (or $H$) gate maps the basis states $\ket{0}$ and $\ket{1}$ to $\frac{1}{\sqrt{2}}(\ket{0} + \ket{1})$ and $\frac{1}{\sqrt{2}}(\ket{0} - \ket{1})$, respectively, creating a superposition with equal amplitudes.
The \emph{controlled-NOT} (CNOT) gate operates on a two-qubit register $\ket{q_0, q_1}$. If the control qubit $\ket{q_0}$ is 1, it inverts the target qubit $\ket{q_1}$, otherwise, the qubits remain the same. Formally, it maps $\ket{q_0, q_1}$ to $\ket{q_0, q_0 \oplus q_1}$.
The \emph{Toffoli gate} (CCNOT) is a universal reversible gate that operates on three qubits. If the first two qubits are both 1, it inverts the third qubit. Otherwise, the qubits remain unchanged. It maps $\ket{q_0, q_1, q_2}$ to $\ket{q_0, q_1, q_0 q_1 \oplus q_2}$.
The \emph{Swap gate} is a two-qubit operator that swaps the states of the two qubits, mapping $\ket{q_0, q_1}$ to $\ket{q_1, q_0}$.

A multi-controlled NOT (M-CNOT) gate, operating on $n$ qubits, flips the target qubit \( |q_{n-1}\rangle \) when all \( n-1 \) control qubits \( |q_i\rangle \) (for \( 0 \leq i < n-2 \)) are set to \( |1\rangle \). Formally, it applies the transformation  
$
|q_0, q_1, \ldots, q_{n-1}\rangle \mapsto |q_0, q_1, \ldots, q_{n-2}, (q_0 \cdot q_1 \cdots q_{n-2}) \oplus q_{n-1}\rangle.
$  
Toffoli (1980) showed that classical multi-controlled gates require ancillary bits~\cite{Toffoli80}, whereas their quantum counterparts avoid this need but require an exponential number of gates, limiting practicality beyond five controls.  
The standard M-CNOT implementation follows Boolean logic and requires \( n-2 \) ancillary qubits, resulting in linear circuit depth~\cite{barenco1995}. Alternative approaches reduce depth to logarithmic~\cite{Yong17,Balauca22} or achieve constant time in specific architectures~\cite{Rasmussen20}.



\smallskip

A key operator in computational models which makes use of a Quantum Random Access Memory (QRAM) is the quantum memory access operator, which enables coherent retrieval of quantum data. Given an address register in superposition, it retrieves the corresponding data while preserving quantum coherence, allowing parallel access to multiple memory locations within a single operation~\cite{Giovannetti2008}.  
A QRAM system consists of a quantum-controlled addressing mechanism and a quantum memory that stores and retrieves data without collapsing superpositions. A common model employs a binary tree structure, where each level encodes an address bit, and quantum-controlled operations traverse the tree coherently.  

Given an address superposition  
$\sum_{j} \alpha_j \ket{j}$,
the QRAM operator $Q$ maps it to a superposition of stored data, formally 
$$Q \sum_{j} \alpha_j \ket{j} \ket{0} = \sum_{j} \alpha_j \ket{j} \ket{D_j},$$ 
where \(\ket{D_j}\) is the data at memory location \( j \). This transformation preserves quantum coherence, enabling efficient quantum data retrieval.  

In principle, the size of a quantum memory register, defined as the number of qubits it comprises, is not inherently limited. However, state of the art QRAM implementations face challenges such as decoherence, noise, scalability limitations, and gate errors. 



\subsection{Adapting Classical Bitwise Operations for Quantum Computation}

Due to the no-cloning theorem, quantum information cannot be copied or arbitrarily deleted; it can only be transferred within the circuit while maintaining reversibility. This fundamental constraint implies that classical logical operations such as AND and OR cannot be directly applied to overwrite the value of a single qubit, as in the classical operations \( (a \leftarrow a \land b) \) or \( (a \leftarrow a \lor b) \), since they are inherently non-reversible. In contrast, the XOR operation preserves reversibility, allowing transformations such as \( b \leftarrow a \oplus b \), which can be implemented using the CNOT gate.  

To effectively manipulate quantum information, a qubit can be conditionally flipped using an \( X \) gate controlled by another qubit, thereby encoding a controlled copy of the state. Moreover, qubit states can be exchanged using the SWAP gate, which ensures that information is transferred without violating unitarity. These operations form the basis for implementing  quantum circuits while respecting the principles of quantum mechanics.  

\begin{itemize}
\item \textbf{Bitwise AND between two registers.} Let $|d\rangle = |d_0 d_1 \ldots d_{n-1}\rangle$ and $|b\rangle = |b_0 b_1 \ldots b_{n-1}\rangle$ be two quantum registers of size $n$.
In light of the above considerations, the operation \( |d_i\rangle \leftarrow |d_i\rangle \land |b_i\rangle \), for $0\leq i<n$, must be implemented using an ancillary qubit \( |a_i\rangle\), initially set to \( |0\rangle \). First, the state of \( |d_i\rangle \) is swapped with \( |a_i\rangle \) to preserve reversibility. Then, a Toffoli gate is applied with \( |a_i\rangle \) and \( |b_i\rangle \) as control qubits and \( |d_i\rangle \) as the target, effectively implementing the operation \( |d_i\rangle \leftarrow |a_i\rangle \land |b_i\rangle \). Since all AND operations can be executed in parallel across the \( n \) qubits, the entire bitwise operation is performed in constant time and requires a register $|a\rangle = |a_0 a_1 \ldots a_{n-1}\rangle$ of \( n \) ancillary qubits.

\item \textbf{Generalized disjunction.} 
The logical OR operation between two qubits \( |b\rangle \) and \( |d\rangle \) stores their disjunction in a third qubit \( |c\rangle \). This can be implemented using a Toffoli gate, which efficiently realizes conjunction, making use of the identity  
$
b \lor d = \neg (\neg b \land \neg d).
$  
As a consequence, the generalized OR operation over the states of an \( n \)-qubit quantum register can be implemented using a multiple-CNOT operator, where the qubits in the register act as controls and a result qubit \( |c\rangle \) serves as the target. This requires negating the qubit states before applying the operator and restoring them to their original values afterward. Additionally, the final state of \( |c\rangle \) must be negated to obtain the correct result.  
The complexity of this operator matches that of the multiple-CNOT, which is \( \mathcal{O}(\log n) \).

\item \textbf{Bitwise shift of a register.} 
In the classical setting, a bitwise shift of an \( n \)-bit register runs in \( \mathcal{O}(\lceil n/w \rceil) \) time but is not reversible, making it unsuitable for quantum computation. Instead, cyclic rotation must be used, preserving all qubit states by permuting them within the register.  
A quantum rotation operator can be implemented in \( \mathcal{O}(\log n) \) time using parallel SWAP gates~\cite{Niroula21}, though a systematic construction for arbitrary register sizes and shifts has been developed only recently~\cite{PavoneV23}. However, Moore and Nilsson~\cite{QuantumCodesMooreNilsson} showed that any permutation, including shifts, can be realized in constant depth, making use of the dihedral group structure. A concrete method for implementing cyclic shifts in constant time has been recently proposed in~\cite{PavoneV24}.

\item \textbf{Incrementing the value of a register.} 
A carry-free increment operator for an \( n \)-qubit quantum register can be implemented using a sequence of \( n \) multiple-CNOT gates. In this sequence, the \( i \)-th multiple-CNOT targets the \( i \)-th least significant qubit, with control qubits ranging from position 0 to \( i-1 \).  
Since a multiple-CNOT gate has a depth of \( \mathcal{O}(\log n) \), the total depth of the increment operator is \( \mathcal{O}(n \log n) \).

\item \textbf{Integer comparison of two registers.} 
A quantum operator for adding two quantum registers can be implemented with linear depth relative to the size of the registers. With a slight modification, the addition operator can be adapted to perform a comparison operation while maintaining the same depth~\cite{Cuccaro04}.

\end{itemize}

\section{Quantum String Matching Algorithms in $\tilde{\mathcal{O}}(n)$}

In this section, we introduce two quantum algorithms that adapt the classical bit-parallel approaches, Shift-And and Shift-Add, to the quantum computing framework. The solutions described in this section do not provide any computational speedup compared to their classical counterparts. However, they demonstrate how the bit-parallelism paradigm can be translated into a quantum algorithm, and how they serve as a basis for the more advanced quantum procedures discussed in the next sections.  

The algorithms described in this section utilize QRAM to access the input text \( y \), on which the search is performed, as well as the table $b$ of size $\sigma$ used to simulate the non-deterministic automaton. Specifically, QRAM\(_y\) is employed to access each of the \( n \) characters composing the text \( y \) via an address register \( |j\rangle \) of size \( \lceil \log(n) \rceil \).  
Furthermore, assuming that the underlying alphabet is ordered and that each of its characters can be mapped to an integer in the range \( [0, \sigma - 1] \), QRAM\(_b\) is used to retrieve the memory registers \( b[c] \) for any $c \in \Sigma$, using a register of size \( \lceil \log(\sigma) \rceil \), containing the value of $c$, to address the information. 

It is assumed that these quantum memory structures are precomputed (ore initialized) before the algorithm is executed. However, it is straightforward to observe that such precomputations can be performed in time \( \tilde{\mathcal{O}}(n) \) and \( \tilde{\mathcal{O}}(m) \), respectively.

\subsection{Quantum Shift-And}

\begin{figure}[t!]
\begin{small}
\textbf{\textsc{Quantum-Shift-And Algorithm:}}\\
\textsc{Input:}\\
-\quad {a \textsc{qram} operator to access a text $y$ of length $n$}\\
-\quad {a \textsc{qram} operator to access the transition vector $B$ of length $\sigma$}\\[0.1cm]
\textsc{Circuit Registers:}\\
-\quad {a register $|a\rangle$ of size $nm$, initialized to $|0\rangle^{nm}$;} (\emph{the ancillary register})\\
-\quad {a register $|b\rangle$ of size $m$, initialized to $|0\rangle^{m}$;} (\emph{the transition vector register})\\
-\quad {a register $|d\rangle$ of size $m$, initialized to $|0\rangle^{m}$;} (\emph{the automaton configuration register})\\
-\quad {a register $|c\rangle$ of size $\lceil \log(\sigma) \rceil$, initialized to $|0\rangle^{\log(\sigma)}$;} (\emph{the character address register})\\
-\quad {a register $|j\rangle$ of size $\lceil \log(n) \rceil$, initialized to $|0\rangle^{\log(n)}$;} (\emph{the shift address register})\\
-\quad {a qubit $|r\rangle$, initialized to $|0\rangle$;} (\emph{the output register})\\[0.1cm]
\textsc{Circuit Procedure:}\\
1.~ \textsc{repeat $n$ times the following sequence}\\
2.~ \qquad \textsc{$|c\rangle \leftarrow$ get a register from QRAM$_y$ with address $|j\rangle$}\\
3.~ \qquad \textsc{$|b\rangle \leftarrow$ get a register from QRAM$_B$ with address $|c\rangle$}\\
4.~ \qquad \textsc{swap $|d_i\rangle$ and $|a_i\rangle$, for $i$ from $0$ to $m-1$ (in parallel)}\\
5.~ \qquad \textsc{perform $|d_{i+1}\rangle \leftarrow |b_i\rangle$ and $|a_i\rangle$, for $i$ from $0$ to $m-2$ (in parallel)}\\
6.~ \qquad \textsc{set $|d_0\rangle$ to $|b_0\rangle$ using the cx operator}\\
7.~ \qquad \textsc{right rotate $|a\rangle$ by $m$ positions}\\
8.~ \qquad \textsc{increment $|j\rangle$}\\
9.~ \textsc{perform $|r\rangle \leftarrow$ generalized-or$(|A[0]_{m-1}\rangle, |A[1]_{m-1}\rangle, \ldots, |A[n-1]_{m-1}\rangle)$}\\
10. \textsc{measure $|r\rangle$ and return the corresponding bit}\\[0.2cm]
\end{small}
    \includegraphics[width=1\linewidth]{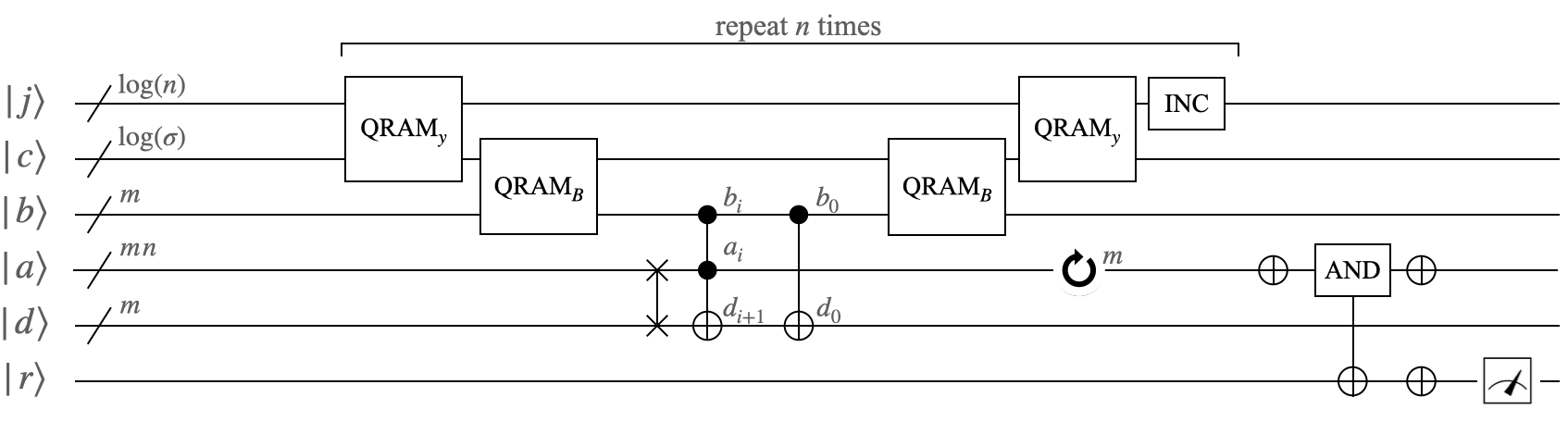}
    \caption{The Quantum Shift-And algorithm, which translates the classical bit-parallel Shift-And algorithm into a quantum framework. (Top) High-level pseudocode describing the procedure executed by the circuit. (Bottom) The quantum circuit implementing the algorithm.}
    \label{fig:circuito-QSA}
\end{figure}

The Quantum Shift-And (\textsc{QSAnd}) algorithm extends the classical SA approach to a quantum computational framework. 
%
%
%
%
%
The algorithm (see Figure \ref{fig:circuito-QSA}) utilizes a quantum register \( |d\rangle \) of \( m \) qubits, initialized to \( |0\rangle^m \), to encode the automaton configurations, where active states in process are represented by qubits set to \( |1\rangle \), while non-active states are set to \( |0\rangle \).

A register \( |b\rangle \) of size \( m \) is used to store transition vector registers during the automaton simulation. Additionally, an ancillary register \( |a\rangle \) of \( nm \) qubits, initialized to \( |0\rangle^{nm} \), is employed to store intermediate computational states. A result qubit \( |r\rangle \), initialized to \( |0\rangle \), is used to detect occurrences of the pattern.  

To address characters in the text of length \( n \), a register \( |j\rangle \) of \( \lceil \log(n) \rceil \) qubits is used to retrieve them via QRAM\(_y\). Similarly, a register \( |c\rangle \) of \( \lceil \log(\sigma) \rceil \) qubits is employed to access the transition vector registers through QRAM\(_b\).


At the beginning of the procedure, the register \( |j\rangle \) is initialized to \( |0\rangle \). The algorithm consists of \( n \) iterations of a fixed sequence of instructions, with \( |j\rangle \) being incremented at the end of each iteration. At the start of each iteration, the register \( |d\rangle \) stores the current automaton configuration, while the first \( m \) qubits of the ancillary register \( |a\rangle \) are set to \( |0\rangle^m \).  

In each iteration, the algorithm executes the following sequence of operations.  

First, the quantum register \( |y[j]\rangle \), representing the current character of the text, is retrieved from QRAM\(_y\) and stored in \( |c\rangle \). The value of \( |c\rangle \) is then used to address QRAM\(_b\), retrieving the quantum register \( |b[y[j]]\rangle \), which encodes the transition information for the current character. This information is stored in the register \( |b\rangle \).  

To correctly simulate the transition, the \( m \) qubits of the automaton register \( |d\rangle \) are swapped with the first \( m \) qubits of the ancillary register \( |a\rangle \), which are in the state \( |0\rangle^m \). This ensures that the previous automaton configuration is preserved.  

Next, a bitwise AND operation is performed in parallel for all \( 0 \leq i < m-1 \) between \( |b_i\rangle \) and \( |a_i\rangle \), updating \( |d_{i+1}\rangle \) accordingly. This operation effectively shifts the automaton states one position to the right, leaving the first state in \( |0\rangle \). It is implemented using a set of parallel Toffoli gates, where \( |b_i\rangle \) and \( |a_i\rangle \) act as controls, and \( |d_{i+1}\rangle \) serves as the target.  

Finally, the first qubit of the automaton configuration, \( |d_0\rangle \), is updated to \( |b_0\rangle \) to simulate the transition from the initial state, which is always active. This is achieved by applying a controlled-\( X \) (CX) gate, using \( |b_0\rangle \) as the control and \( |d_0\rangle \) as the target.  

To ensure proper automaton evolution, the register \( |a\rangle \) is rotated by \( m \) positions, resetting the qubits involved in the next transition to \( |0\rangle \).  Additionally, to ensure the reversibility of this procedure, it is necessary to restore the values of the involved registers for the next iteration. As previously discussed, this is achieved through an uncompute operation, which in this context corresponds to the application of the two QRAM access operators.

We observe that this process preserves the \( n \) different configurations of the automaton within the auxiliary register \( |a\rangle \), whose total size is \( nm \). By decomposing this register into \( n \) sub-registers of size \( m \), denoted as \( |A_i\rangle \) for \( 0 \leq i < n \), each sub-register \( |A_i\rangle \) contains the automaton configuration after the \( i \)-th iteration of the procedure.  
Since the final state of each configuration determines whether the pattern occurs, we can conclude that the pattern \( x \) appears at position \( j \) in the text \( y \) if the \( m \)-th qubit of register \( |A_j\rangle \) is in the state \( |1\rangle \).  

Thus, once the entire text has been processed (i.e. at the end of the $n$ iterations), a generalized OR is performed between the $m$-th qubits of the registers \( |A_i\rangle \) for \( 0 \leq i < n \) and the result id stored in \( |r\rangle \). Finally, the register $|r\rangle$ is measured. If the measurement outcome is \(|1\rangle\), it confirms at least one occurrence of the pattern within the text.

\smallskip

To analyze the asymptotic time complexity of the algorithm in terms of quantum circuit depth, we make the following observations.  
The QRAM access operations at lines 2 and 3 have a depth of \( \mathcal{O}(\log n) \) and \( \mathcal{O}(\log \sigma) \), respectively. The swap and bitwise AND operations at lines 4 and 5 are executed in parallel across all qubits of the quantum register, resulting in constant depth. Similarly, the rotation operation at line 7 can be performed with constant depth, whereas the increment operation on register \( |j\rangle \) has a complexity of \( \mathcal{O}( \log m) \).  

Finally, the generalized OR operation over the \( n \) elements of register \( |a\rangle \) has a depth of \( \mathcal{O}(\log n) \). Therefore, the overall depth complexity of the algorithm for solving formulation (1) of the problem is \( \tilde{\mathcal{O}}(n) \).  
In addition, it is straightforward to see that problem formulations (2) and (3) can also be solved in \( \tilde{\mathcal{O}}(n) \) time. However, we do not elaborate on these details due to space constraints and their limited relevance to the following discussion.


\subsection{Quantum Shift-Add}

\begin{figure}[t!]
\begin{small}
\textbf{\textsc{Quantum-Shift-Add Algorithm:}}\\
\textsc{Input:}\\
-\quad {a QRAM operator to access a text $y$ of length $n$}\\
-\quad {a QRAM operator to access the transition vector $B$ of length $\sigma$}\\[0.1cm]
\textsc{Circuit Registers:}\\
-\quad {a register $|a\rangle$ of size $\lceil \log(m) \rceil$, initialized to $|k\rangle$;} (\emph{the bound register})\\
-\quad {a register $|b\rangle$ of size $m$, initialized to $|0\rangle^{m}$;} (\emph{the transition vector register})\\
-\quad {a register $|d\rangle$ of size $n\lceil \log(m) \rceil$, initialized to $|0\rangle^{n\log(m)}$;} (\emph{the automaton configuration register})\\
-\quad {a register $|c\rangle$ of size $\lceil \log(\sigma) \rceil$, initialized to $|0\rangle^{\log(\sigma)}$;} (\emph{the character address register})\\
-\quad {a register $|j\rangle$ of size $\lceil \log(n) \rceil$, initialized to $|0\rangle^{\log(n)}$;} (\emph{the shift address register})\\
-\quad {a register $|s\rangle$ of size $n$, initialized to $|0\rangle^{n}$;} (\emph{the shift occurrence register})\\
-\quad {a qubit $|r\rangle$, initialized to $|0\rangle$;} (\emph{the output register})\\[0.1cm]
\textsc{Circuit Procedure:}\\
1.~ \textsc{repeat $n$ times the following sequence}\\
2.~ \qquad \textsc{$|c\rangle \leftarrow$ get the register from QRAM$_y$ with address $|j\rangle$}\\
3.~ \qquad \textsc{$|b\rangle \leftarrow$ get the register from QRAM$_B$ with address $|c\rangle$}\\
4.~ \qquad \textsc{increment $|D_i\rangle$ if $|b_i\rangle = |1\rangle$, for $i$ from $0$ to $m-1$ (in parallel)}\\
5.~ \qquad \textsc{set $|s_{m-1}\rangle = |1\rangle$ if $|D_{m-1}\rangle \geq |a\rangle$}\\
6.~ \qquad \textsc{right rotate $|d\rangle$ by $\lceil \log(m) \rceil$ position}\\
7.~ \qquad \textsc{right rotate $|s\rangle$ by $1$ position}\\
8.~ \qquad \textsc{increment $|j\rangle$}\\
9.~ \textsc{perform $|r\rangle \leftarrow$ generslized-or$(|r_0\rangle, |r_1\rangle, \ldots, |r_{n-1}\rangle)$}\\
10. \textsc{measure $|r\rangle$ and return the corresponding bit}\\[0.2cm]
\end{small}
    \includegraphics[width=1\linewidth]{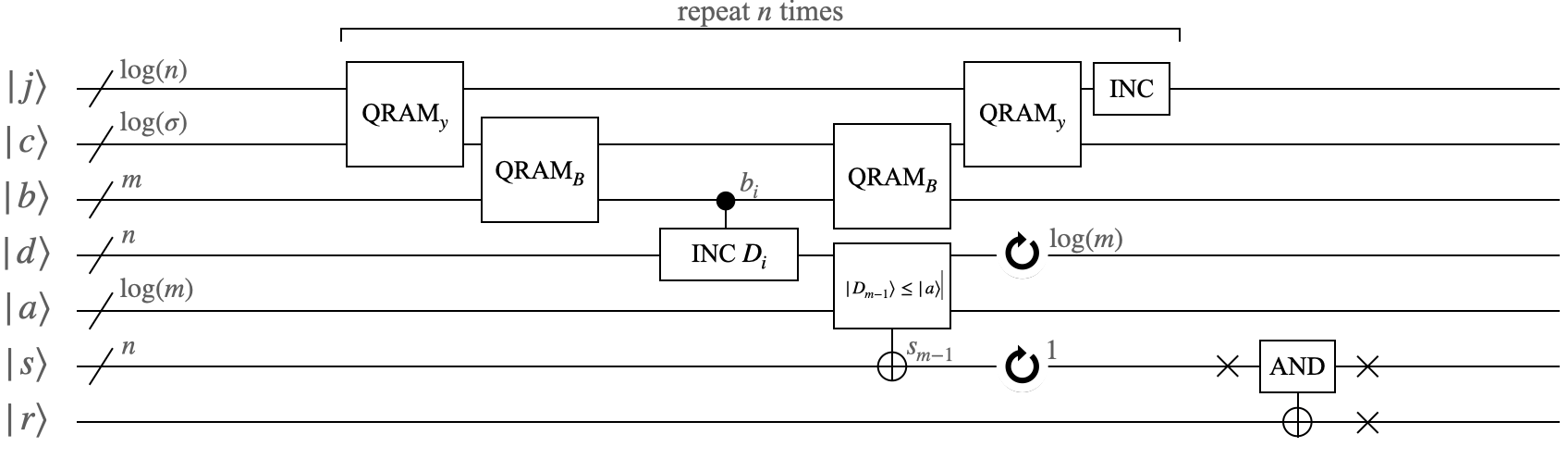}
    \caption{The Quantum Shift-Add algorithm, which translates the classical bit-parallel Shift-Add algorithm into a quantum framework. (Top) High-level pseudocode describing the procedure executed by the circuit. (Bottom) The quantum circuit implementing the algorithm.}
    \label{fig:circuito-QSAdd}
\end{figure}

The Quantum Shift-Add (\textsc{QSAdd}) algorithm extends the classical bit-parallel Shift-Add method to a quantum computational framework, enabling  approximate string matching with at most \( k \) errors. Unlike the \textsc{QSAnd} algorithm, which uses a single qubit per position to track state transitions, \textsc{QSAdd} requires \(\lceil \log m \rceil\) qubits per position to count mismatches dynamically. This allows the algorithm to determine whether the number of mismatches remains within the allowed threshold $k$ during execution.  

Along the same lines of \textsc{QSAnd}, The \textsc{QSAdd} algorithm (depicted in Figure~\ref{fig:circuito-QSAdd}) begins with the initialization of multiple quantum registers that store the state of the automaton $|d\rangle$, the transition vectors $|b\rangle$, and shift occurrence register $|s\rangle$. The automaton register $|d\rangle$ has size $m\lceil \log(m)\rceil$ and we assume it is the concatenation of $n$ sub-registers of size $\lceil \log(m)\rceil$, which we refer as $D_i$, for $0\leq i <n$.  A bound register $|a\rangle$, of size $\lceil \log(m)\rceil$, is initialized with the value of the bound $k$ and is used to detect approximate occurrences of the pattern. The algorithm works iteratively, processing one character of the text at a time until the entire input string has been examined. At each iteration, the algorithm retrieves the current character $|c\rangle$ from the text using a QRAM lookup and maps it to a corresponding transition vector $|b\rangle$, which encodes the state transitions of the automaton.

Once the transition vector $|b\rangle$ is loaded, the automaton’s internal state $|d\rangle$ is updated. This is achieved through a parallelized increment operation, which modifies the configuration of the automaton whenever a transition is valid. A crucial step in this process is checking whether the automaton has reached a final accepting state, i.e. if $|D_{m-1}\rangle \leq |a\rangle$, which would indicate a successful match. If this condition is met, the corresponding position in the shift register $|s\rangle$ is marked.
After processing the character, the automaton state $|d\rangle$ is shifted to the right of $\lceil \log(m)\rceil$ positions, effectively propagating its configuration for the next step. Similarly, the shift occurrence register is rotated, ensuring that valid match positions are updated consistently throughout the execution. The text index register $|j\rangle$ is then incremented, advancing the scanning process.
Once all characters in the text have been processed, the algorithm performs a final quantum OR operation across the shift register $|s\rangle$, collapsing the information into a single output qubit $|r\rangle$. This final step determines whether at least one match has been found in the text. A measurement of this qubit provides the final result, indicating the presence or absence of the pattern in the text.
 
\smallskip

The time complexity of the algorithm can be analyzed by considering the sequence of operations performed at each iteration. Since the main procedure consists of $n$ iterations, the key computational costs arise from the fundamental steps executed in each cycle.  

First, the algorithm retrieves data from QRAM, an operation that incurs a cost of $\mathcal{O}(\log n)$ for accessing the text and $\mathcal{O}(\log m)$ for retrieving the transition vector. The automaton update involves $m$ controlled increments applied in parallel to registers of size $\log m$, contributing an overall cost of $\mathcal{O}(\log m)$ per iteration. Additionally, the comparison between the register $|D_{m-1}\rangle$ and the bound register $|a\rangle$, both of size $\log m$, introduces another $\mathcal{O}(\log m)$ operation.  
Once all $n$ iterations are completed, a generalized quantum OR operation is applied to aggregate the results across $n$ qubits, which adds a cost of $\mathcal{O}(\log n)$.  
As a result, the total time complexity, in terms of circuit depth, is given by $\mathcal{O}(n \log m + \log n)$ to solve formulation (1) of the problem. This simplifies to $\tilde{\mathcal{O}}(n)$, indicating that the algorithm is near-linear.


\section{Enhancing Quantum String Matching via Grover’s Search}

In the previous section, we demonstrated how a quantum algorithm can be implemented to verify the presence of a pattern \( x \) of length \( m \) within a text \( y \) of length \( n \) in time \( \tilde{\mathcal{O}}(n) \). This result, which addresses formulation (1) of the string matching problem, is achieved by directly translating a bit-parallelism-based algorithm into the quantum computing framework.

In this section, we use this operator as the base for developing a new quantum algorithm that solves the same problem in time \( \tilde{\mathcal{O}}(m\sqrt{n}) \). We also demonstrate how this approach can be further refined to achieve a complexity of \( \tilde{\mathcal{O}}(\sqrt{n}) \). Such solutions make use of the well-known unstructured search algorithm, introduced by Grover in 1996.

\smallskip

Specifically, Grover’s algorithm is a fundamental quantum search technique that locates a unique element within an unstructured dataset of size \( n \) in \( \mathcal{O}(\sqrt{n}) \) operations~\cite{Grover96}, providing a quadratic speedup over classical brute-force methods.  
%
%
Formally, the algorithm addresses the problem of identifying a unique solution \( w \) to a given function \( f: \{0,1\}^{\log(n)} \to \{0,1\} \), where \( f(w) = 1 \) for exactly one input \( w \). 

The core idea is to iteratively amplify 
the probability of obtaining the correct solution through iterative quantum rotations. It starts with an equal superposition of all inputs and incrementally rotates the quantum state towards the target state, where the solution \( w \) has amplitude 1.  
Each iteration consists of two main operations: the Phase Oracle \( U_f \), which flips the amplitude of \( w \), and the Grover Diffusion Operator, which reflects the state around the uniform superposition \( |+\rangle^{\otimes n} \), reinforcing the amplitude of \( w \). 
The total rotation per iteration is approximately \( 2 / \sqrt{n} \), requiring about \( \frac{\pi}{4} \sqrt{n} \) iterations to maximize the probability of measuring \( w \).  
From a computational complexity perspective, each iteration consists of one application of the phase oracle and one application of the diffusion operator. Since the latter requires a multi-controlled \( Z \) gate acting on \( n \) qubits, its circuit depth scales as \( \mathcal{O}(\log n) \). Therefore, the overall complexity of Grover’s search is given by \( \mathcal{O}(\sqrt{n}(T(n) + \log n)) \), where \( T(n) \) denotes the time complexity of implementing the oracle. 

The algorithm generalizes to cases where \( f \) has multiple valid solutions. If there are \( r \) solutions, the number of required iterations remains \( \mathcal{O}(\sqrt{n}) \) but is more precisely given by \( \frac{\pi}{4} \sqrt{n / r} \)~\cite{Brassard2002}, which is known to be optimal~\cite{Brassard98}. However, when \( r \) is unknown, it must be estimated, often requiring techniques based on Quantum Phase Estimation. Finally, if the goal is to find all \( r \) solutions, the number of required queries is \( \Theta(\sqrt{n r}) \)~\cite{Ambainis04}.

\subsection{A First Jump From $\tilde{\mathcal{O}}(n)$ to $\tilde{\mathcal{O}}(m\sqrt{n})$}

We observe that the string matching problem can be reformulated as a quantum database search problem, where the objective is to determine whether a given pattern appears within a set of substrings of the text. This reformulation enables the use of Grover’s algorithm to solve the problem, making use of quantum parallelism to reduce the computational complexity.  
  
The resulting algorithm, whose high level pseudocode is schematized in Figure \ref{code:QSA-G}, operates by dividing the text into \( n-m \) overlapping blocks, each of size \( m \), where \( m \) corresponds to the length of the pattern, and where each block shares $m-1$ characters with the next block. With this text partitioned, a quantum search is performed in parallel across all blocks to determine whether the pattern is present in any of them. If the search identifies at least one block containing the pattern, the corresponding index is returned, solving problem (1). 

\begin{figure*}[!t]
\begin{small}
\begin{quote}
\textbf{\textsc{Quantum Procedure A:}}\\[0.1cm]
1. \textsc{divide the text into $n-m$ overlapping blocks of size $m$;}\\
2. \textsc{set $|j\rangle$ as the superposition of all value between $|0\rangle$ and $|n\rangle$;}\\
3. \textsc{repeat $\pi/4 \sqrt{n}$ times the following sequence:}\\
4. \qquad \textsc{run the QSAnd algorithm on the block beginning at position $|j\rangle$}\\
5. \qquad \textsc{execute the Grover Diffuser operator}\\
6. \textsc{Measure $|j\rangle$ to identify a shift $i$ of the text where the pattern occurs;}\\
7. \textsc{Return $i$}
\end{quote}
\end{small}
\caption{\label{code:QSA-G}The Quantum String Matching algorithm which makes use of the Grover unstructured search, working in $\tilde{\mathcal{O}}(m\sqrt{n})$ time.}
\end{figure*}

The key advantage of this method lies in the quadratic speedup provided by Grover’s algorithm. 
Using \textsc{QSAnd} or \textsc{QSAdd}, verifying the presence of an occurrence of the pattern within a single block of length \( m \) requires \( \tilde{\mathcal{O}}(m) \) operations and $\mathcal{O}(m^2)$ auxiliary qubits. 
However, the algorithm executes the Grover search over the set of substrings performing \( \mathcal{O}(\sqrt{n}) \) queries. Since each search requires $\tilde{\mathcal{O}}(m)$ time, the total complexity of the quantum algorithm is \( \tilde{\mathcal{O}}(m \sqrt{n}) \), representing a significant improvement over the classical methods. 


\subsection{A Final Shift From $\tilde{\mathcal{O}}(m\sqrt{n})$ to $\tilde{\mathcal{O}}(\sqrt{n})$}


%
In this section, we present our final solution to the problem, which aims to reduce the search space of Grover’s algorithm, thereby decreasing the number of iterations required to amplify the result vector. This reduction is achieved by partitioning the text into blocks larger than \( m \), leading to a lower number of substrings to be searched.  

However, while this approach reduces the search space, it also increases the size of the blocks on which linear-time algorithms must be applied during each iteration, introducing a trade-off between search space reduction and computational overhead per iteration.  

\begin{figure*}[!t]
\begin{small}
\begin{quote}
\textbf{\textsc{Quantum Procedure B:}}\\[0.1cm] 
1. \textsc{divide the text into $n/(K-m+1)$ overlapping blocks of size $K$;}\\
2. \textsc{set $|j\rangle$ as the superposition of all value between $|0\rangle$ and $|n/K\rangle$;}\\
3. \textsc{repeat $\pi/4 \sqrt{n/K}$ times the following sequence:}\\
4. \qquad \textsc{run the QSAand algorithm on the $j$-th block of the text}\\
5. \qquad \textsc{execute the Grover Diffuser operator}\\
6. \textsc{Measure $|j\rangle$ to identify a block $i$ of the text where the pattern occurs;}\\
7. \textsc{Use \textbf{Quantum Procedure A} to search in the $i$-th block for a valid shift $h$}\\
8. \textsc{Return $Ki + h$}
\end{quote}
\end{small}
\caption{\label{code:QSA-G2}The Quantum String Matching algorithm which makes use of the Grover unstructured search, working in $\tilde{\mathcal{O}}(\sqrt{n})$ time.}
\end{figure*}

Formally, let $K$ be the size of each block into which the text is partitioned, with $K > m$. The number of blocks to search in the text will be $n/(K - m + 1)$, as adjacent blocks must share $m - 1$ characters to avoid skipping valid shifts. However, for $K$ significantly larger than $m$, this value approximates to $n/K$. 
Thus, Grover's algorithm will search a space of $n/K$ elements, requiring $\tilde{\mathcal{O}}(\sqrt{n/K})$ iterations. Each iteration now takes $\mathcal{O}(K + \log(K))$ time, so the overall depth of the circuit implementing the algorithm will be $\tilde{\mathcal{O}}((K+\log(K)) \sqrt{n/K}) = \tilde{\mathcal{O}}(\sqrt{nK})$.
Therefore, by partitioning the text into segments of size $K = \log(n)$, we achieve a depth of $\tilde{\mathcal{O}}(\sqrt{n} \log^{1/2}(n)) = \tilde{\mathcal{O}}(\sqrt{n})$.

Observe that, in this context, each iteration of the search algorithm checks for the presence of the pattern within blocks larger than $m$, allowing the possibility of multiple occurrences within the same block.  
This is consistent with the resolution of case (1) of the problem.


Thus, if Grover's search algorithm can identify the index $i$ of a  block where the pattern occurs, determining the exact position of the pattern within the text requires executing the quantum string matching algorithm on such block using $K = m$. This step will return the offset $j$ within the block where the pattern is found. This will allow us to state that the pattern occurs at position $Ki + j$ in the text.
The depth of the circuit for this additional refined search is $\mathcal{\tilde{\mathcal{O}}}(m \log^{1/2}(n))$, ensuring that the overall depth remains $\tilde{\mathcal{O}}(\sqrt{n})$.

\section{Final Discussion}

In this work, we have shown how a string matching algorithm based on bit-parallelism can be efficiently translated into the quantum computing framework while preserving its computational complexity, up to a logarithmic factor, in terms of circuit depth. Additionally, by making of Grover’s search algorithm, we have developed a quantum solution that solves formulation (1) of the problem in \( \tilde{\mathcal{O}}(\sqrt{n}) \) time.  
Although we have omitted implementation details due to space constraints, we observe that formulation~(2) could be addressed using Quantum Phase Estimation applied to the bit-parallelism-based operator. Similarly, for formulation~(3), when \( r \) solutions are present, Grover’s algorithm requires \( \mathcal{O}(\sqrt{nr}) \) iterations to identify all occurrences, leading to a complexity of \( \tilde{\mathcal{O}}(\sqrt{nr}) \), which provides a quantum advantage when \( r = o(n) \).  
While similar complexity results exist in the literature, our approach offers a systematic way to translate bit-parallelism-based algorithms into the quantum framework, demonstrating that a broad class of classical string matching techniques can benefit from quantum parallelism, achieving a quadratic speedup.  
Moreover, its simplicity allows direct implementation as a quantum circuit. Given the extensive research on bit-parallelism techniques for non-standard string matching problems, our results pave the way for further studies and applications, extending quantum speedups to a broader range of text-searching challenges.

\end{document}